\documentclass{article}
\usepackage{amsmath,amsfonts,amsthm}
\usepackage{graphicx}
\usepackage{xcolor}
\usepackage{fullpage}
\usepackage{lscape}

\newtheorem{theorem}{Theorem}[section]

\theoremstyle{definition}
\newtheorem{definition}[theorem]{Definition}
\newtheorem{example}[theorem]{Example}
\newtheorem{question}[theorem]{Question}

\author{
  Seung Hyeon Mandy Hong\\
  \texttt{hong\_m1@denison.edu}
  \and
  May Mei\\
  \texttt{meim@denison.edu}
}
\title{The Game of Life on the Robinson Triangle Penrose Tiling: Still Life}

\begin{document}
\maketitle

\begin{abstract}
We investigate Conway's Game of Life played on the Robinson triangle Penrose tiling. In this paper, we classify all four-cell still lifes.
\end{abstract}

\section{Introduction}

John von Neumann is often credited with introducing cellular automata in \cite{vN51}, while Martin Gardner popularized John Conway's Game of Life in \cite{Gar70}. Notably, three simple rules on a square lattice yield remarkably complex behavior. Indeed \cite{Gar70} introduces a pentomino (five cell configuration) whose behavior had not stabilized after 460 generations. Martin Gardner is also credited with bringing attention to the Penrose tiling in \cite{Gar77}, which incidentally Conway also contributed to. Here, following the investigations begun in \cite{OS10a} and \cite{OS10b}, we explore the intersection of these two topics by playing Game of Life on the Robinson triangle variation of the Penrose tiling.

\section{Penrose Tiling}

Following \cite{GS16} we introduce some definitions about tilings in the most general sense. See also $\cite{Ada22}$ for an accessible introduction.
\begin{definition}
A \emph{tiling} $\mathcal{T}=\{T_i: i=1,2,3,\dots\}$ is a family of closed sets called \emph{tiles}  so that
\begin{enumerate}
\item $T_i$ is topologically equivalent to a closed disk,
\item the union of tiles is the whole plane, or $\bigcup _{i \in \mathbb{N}} T_i = \mathbb{R}^2$,
\item the interiors of tiles are pairwise disjoint, or $T_i^\circ \cap T_j^\circ=\emptyset$ for $i \neq j$.
\end{enumerate}
\end{definition}
A collection of subsets that satisfy condition 2 above is said to be a \emph{covering} and one that satisfies condition 3 above is said to be a \emph{packing}. Thus, a tiling is a family of closed sets that is both a covering and a packing. In many settings we impose additional restrictions such as tiles are compact, have finite volume, have a boundary that is a Jordan curve, etc. In this paper, all tiles are polygonal so we refer additionally to \emph{vertices} and \emph{edges} and require that polygons meet \emph{edge-to-edge}.
\begin{definition}
We say that two (distinct) tiles are \emph{adjacent} if they have an edge in common and two (distinct) tiles are \emph{neighbors} if their intersection is nonempty. We say that the \emph{neighborhood} of a tile to be the collection of its neighbors.
\end{definition}
\noindent Note, this corresponds to the \emph{Moore neighborhood}. In the context of tilings, this is usually referred to as the \emph{first corona}.

Following Chapters 5 and 6 of \cite{BG13}, we briefly discuss the construction of a tiling using \emph{substitution} or \emph{inflation and subdivision}. See also \cite{Fra08} and the references contained therein as an introduction to the topic.

\begin{definition}
We say that the equivalence class of a tiling $\mathcal{T}$ (with respect to translation in $\mathbb{R}^2$) is a \emph{protoset}. That is, every element of $\mathcal{T}$ can be obtained by translating an element of $\mathcal{P}=\{T_1, \dots , T_n\}$ and no element of $\mathcal{P}$ can be obtained by translating a different element of $\mathcal{P}$. Often, we restrict ourselves to finite protosets. Elements of the protoset are referred to as \emph{prototiles}.
\end{definition}

\begin{definition}
Consider a finite protoset $\mathcal{P}=\{T_1,\dots,T_n\}$. An \emph{inflation rule} or \emph{substitution rule} with inflation factor $\lambda>1$ is a map
$$T_i \mapsto \lambda T_i \mapsto \bigcup_{j=1}^n (T_j + A_{ji})$$
where $A_{ji} \in \mathbb{R}^2$ are chosen to ensure that the set on the right is a packing. Stated simply, inflate tile $T_i$ by a factor of $\lambda$, and subdivide $\lambda T_i$ into tiles.
\end{definition}

\begin{example}
The square lattice can be thought of as a substitution tiling with a single prototile (the unit square) and an inflation factor $\lambda=2$.
\end{example}

\begin{figure}
\includegraphics[width=\textwidth]{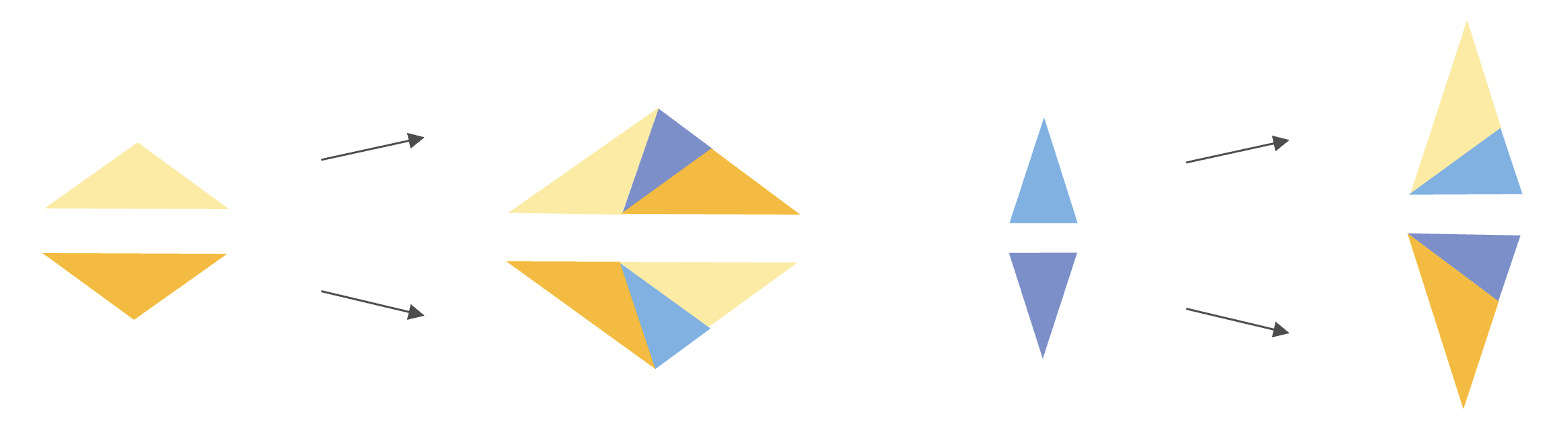}
\caption{The Robinson triangle substitution rule.}\label{fig:triangle}
\end{figure}

\begin{example}
Following \cite{Bie}, the tiling we discuss in this paper is the Robinson triangle variation of the Penrose tiling, pictured in Figure~\ref{fig:triangle} which has an inflation factor of $\lambda=\displaystyle\frac{\sqrt{5}+1}{2}$, the golden ratio.
\end{example}

\section{The Game of Life and Neighborhoods}\label{sec:GoL}
While many variations exist, we will use the classical rules for the Game of Life (GoL) as presented in \cite{Gar70}. All tiles (or cells) exist in two states (alive or dead) and evolve under the following rules:
\begin{enumerate}
\item A live tile remains alive if exactly two or three of its neighbors are alive.
\item A live tile dies if fewer than two or more than three of its neighbors are alive.
\item A dead tile comes alive if exactly three of its neighbors are alive.
\end{enumerate}
As is clear from this description, the structure of neighborhoods is deeply consequential as we consider the evolution of states in the GoL. Classical GoL is played on a square lattice, so each tile has an identical neighbor. The aperiodicity of the Penrose tiling results in a variety of neighborhoods.

\begin{figure}
\includegraphics[width=\textwidth]{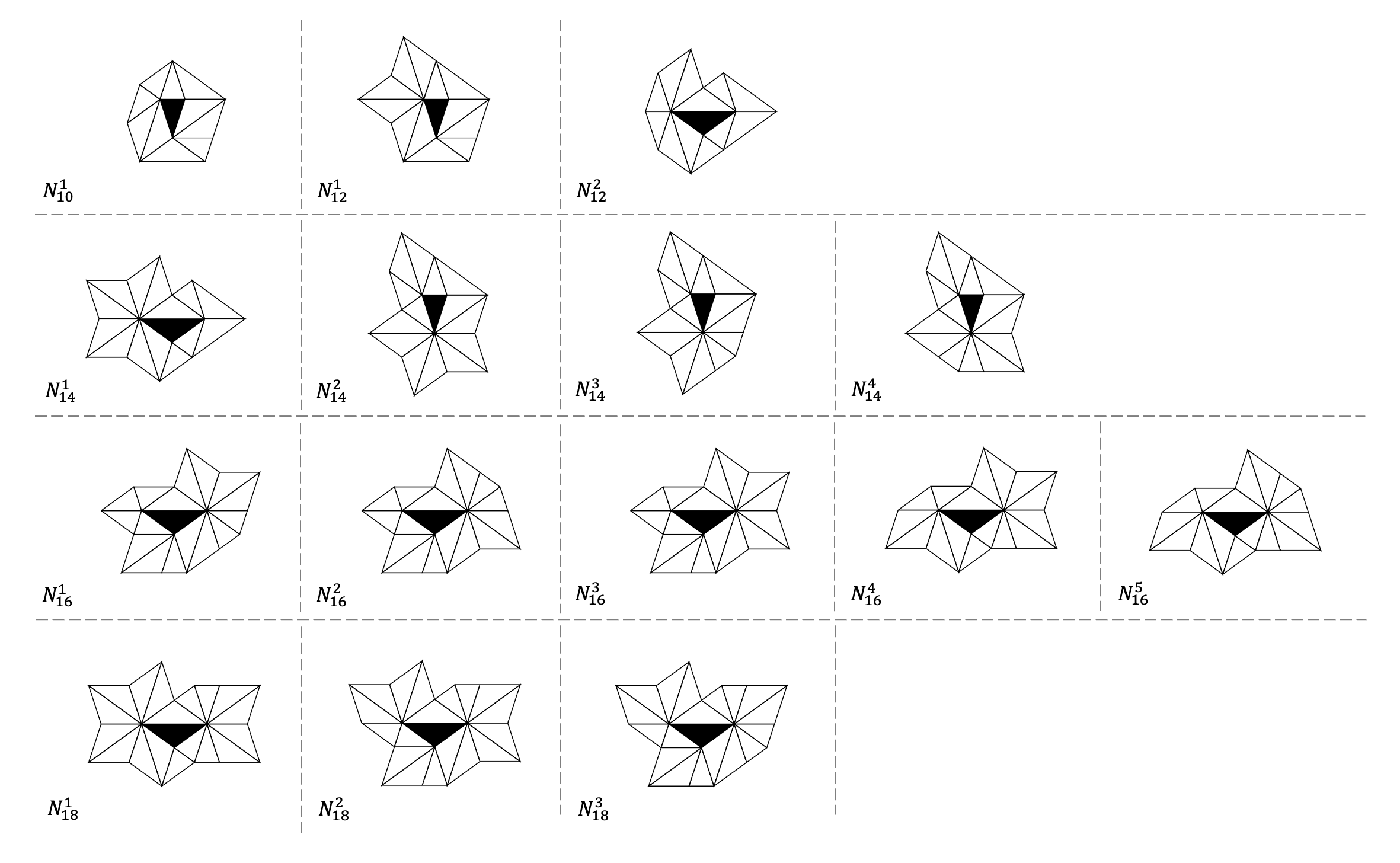}
\caption{All neighborhoods of the Robinson triangle Penrose tiling.}\label{fig:neighborhoods}
\end{figure}

We classify all neighborhoods of this Penrose tiling in Figure~\ref{fig:neighborhoods}. While it is standard to think of the substitution as being define on four prototiles, there are two non-congruent shapes. As a convention, we orient the center tile of the neighborhood pointing downwards. Thus, neighborhoods also appear reflected across a horizontal axis or rotated by $180^\circ$.

See Chapter 5.2 of \cite{BG13} for a discussion of local derivability in general and Chapter 10.3 of \cite{GS16} for specific derivations for different Penrose tilings specifically. The neighborhoods in Figure~\ref{fig:neighborhoods} can be obtained by applying local derivability rules to the rhombus Penrose tiling as seen in Fig.18.12 in \cite{OS10b} or the kite-dart Penrose tiling as seen in Figure 10.5.3 in \cite{GS16}.

\section{Still Lifes}

Having identified all neighborhoods of the Robinson triangle Penrose tiling, we now proceed to the main result of this paper: an algorithm to classify all four-cell still lifes. In classical GoL, there are two such still lifes: the block and the tub.

It is clear that there can be no one-cell or two-cell still lifes. A three-cell configuration can only be a still life if each cell has exactly two neighbors, that is, all cells are pairwise neighbors. But one may verify that all such configurations involve a dead cell with three live neighbors, which results in a birth. Thus, there are no three-cell still lifes.

\begin{figure}
\begin{center}
\includegraphics[width=0.3\textwidth]{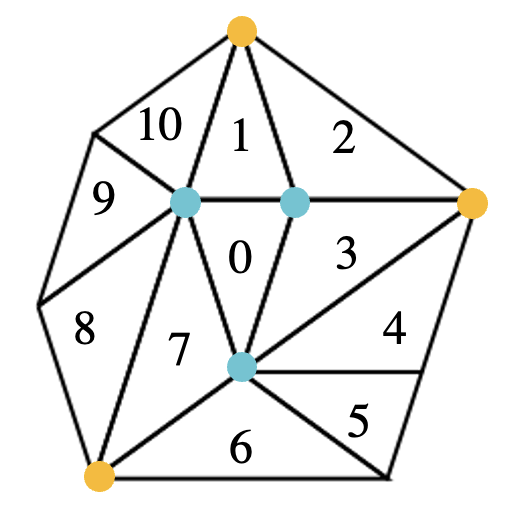}
\end{center}
\vspace{-0.2in}
\caption{There are three inside vertex groups in the $N_{10}^1$ neighborhood: $\{0,1,2,3\}$, $\{0,3,4,5,6,7\}$, and $\{0,1,7,8,9,10\}.$ The outside vertex groups with three or more tiles are $\{1,2,10\}$, $\{2,3,4\}$, and $\{6,7,8\}$. } \label{fig:insideoutside}
\end{figure}

Since we have classified all neighborhoods in Figure~\ref{fig:neighborhoods}, we may classify all four-cell still lifes. There are 15 neighborhoods with a median of 16 neighbors around the center cell, we could brute force approximately $15\cdot\binom{17}{4}$ configurations. Instead, we have optimized by classifying all configurations in which a birth or death may occur and removing them from consideration. One may inspect all remaining configurations and verify that they are indeed still lifes. The following implementation can be found at \cite{Hon23}. First, we begin with the assumption that the center cell (the shaded cell in Figure~\ref{fig:neighborhoods}) and three neighboring cells are alive. We say that the \emph{inside vertex group} is a collection of tiles that share the same vertex of the center tile. There are three (not necessarily disjoint) inside vertex groups. We say that the \emph{outside vertex group} is a collection of tiles that share the same vertex on the boundary of the neighborhood. Refer to Figure~\ref{fig:insideoutside}. A configuration is a still life just in case there are no births and no deaths in the next generation. As the center cell will always have three neighbors, no deaths occur. Thus, we algorithmically remove all configurations that may result in a birth. 

\begin{figure}
\includegraphics[width=\textwidth]{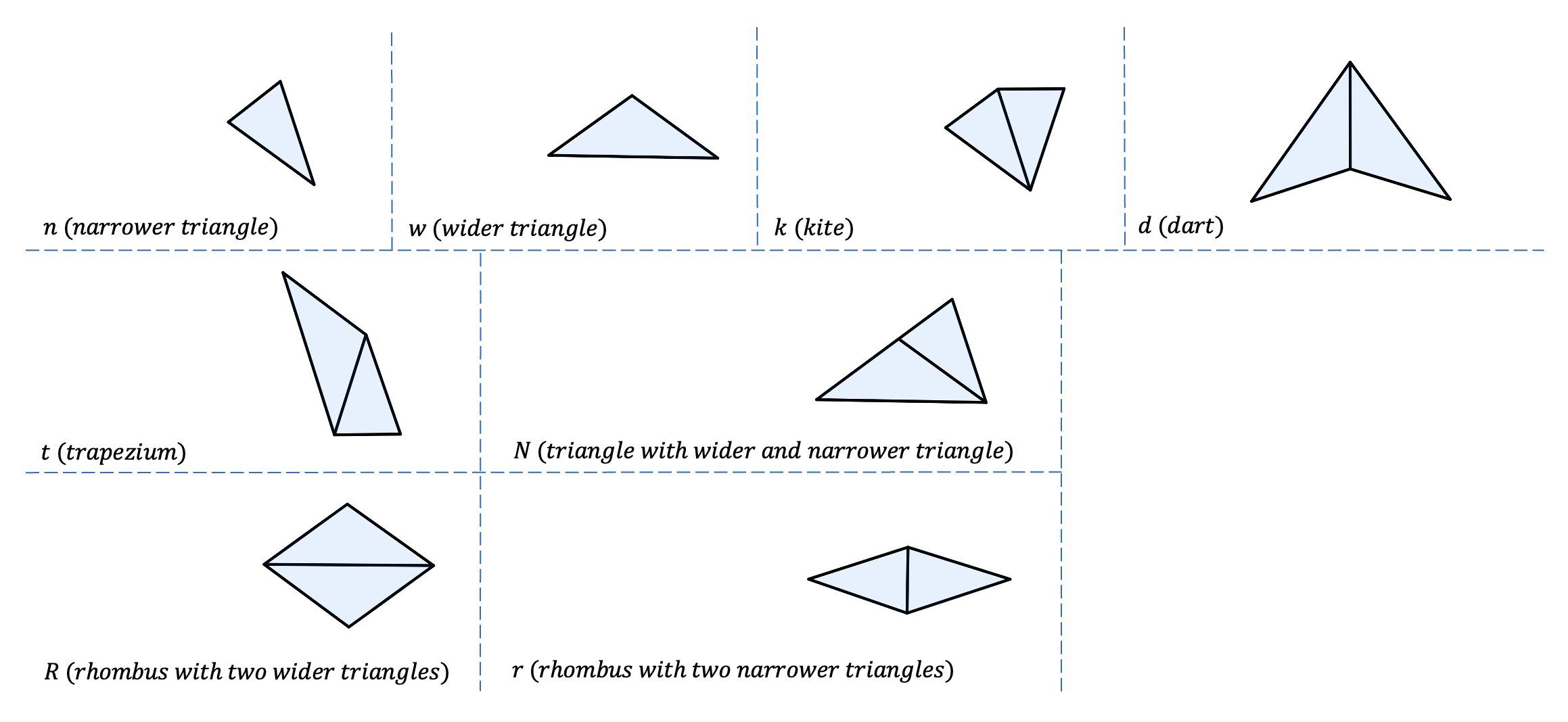}
\caption{Notation for one- and two-cell configurations that we use in Figure~\ref{fig:classification}.}\label{fig:key}
\end{figure}

\begin{figure}
\includegraphics[width=\textwidth]{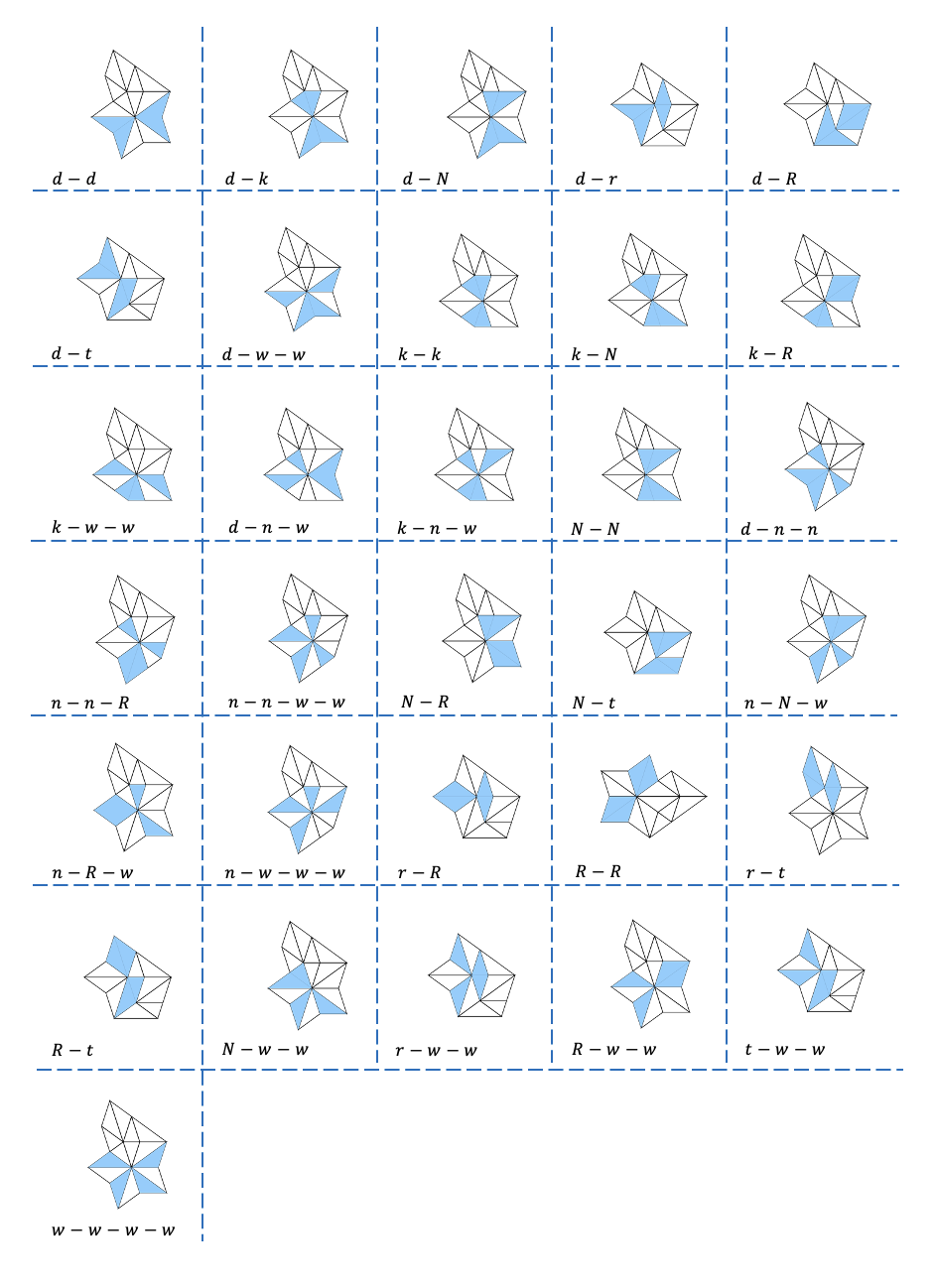}
\caption{Four-cell still lifes in the Robinson triangle Penrose Game of Life.}\label{fig:classification}
\end{figure}

\begin{enumerate}
\item First, we remove all configurations that three live cells share a at least one neighbor outside of the neighborhood. This would result in a birth. Note, as the tiling meets edge-to-edge, there is necessarily a tile adjacent to each edge on the boundary.
\item Next, we consider each of the three inside vertex groups. If precisely three tiles are adjacent to this vertex along with a fourth cell in a different inside vertex group that is not the center cell, a birth occurs. We remove all such configurations.
\item Next, we consider each of the outside vertices. As the center tile is alive, tiles can share a vertex with both the center tile and a boundary vertex. Thus, if two tiles in any outside vertex group and the fourth tile that is not related to the unselected tile from the same outside vertex group are alive, a birth occurs. We remove all such configurations.

\item Finally, we consider configurations where birth occurs in a tile that belongs to two inside vertex groups. We remove all such configurations.
\end{enumerate}
Now, we continue with the assumption that the center cell is dead and four neighboring cells are alive. Steps 1--4 proceed as above with a modification to Step 3, as the center tile is dead. Instead of the center tile, we select one tile that is related to the unselected tile from the chosen outside vertex group as alive. But now, we have a new possibility to contend with: death. 
\begin{enumerate} \setcounter{enumi}{4}
\item If a live cell has 0 or 1 live neighbors, it dies at the next generation. We remove all such configurations.
\end{enumerate}
We present our classification of all four-cell still lifes in Figure~\ref{fig:classification} using the notation introduced in Figure~\ref{fig:key} in Tables 1\textendash4.

\section{Questions}

In Section~\ref{sec:GoL}, we state the classical GoL rules in which every cell has precisely 8 neighbors. Thus, Rule 1 states that births occur when slightly fewer than half of a cell's neighbors are alive. However, in this variation of GoL, cells have 10--18 neighbors. It is then interesting to consider the case that the birth threshold is changed.

\begin{question}
How does this result, i.e. the pattern of still lifes, depend on the threshold for births and deaths?
\end{question}

As still lifes are, by definition, stationary, they lend themselves well to a classification scheme. Two natural questions that follow are then:

\begin{question}
Do any gliders exist in the Robinson triangle Penrose Game of Life?
\end{question}
In \cite{Gou12}, a glider in the rhombus variation of the Penrose Game of Life was presented. As the Robinson triangle variation is easily derivable from the rhombus, it seems natural to ask whether this tiling admits gliders. In our random simulations, we did not find any gliders.

\begin{figure}[h]
\includegraphics[width=\textwidth]{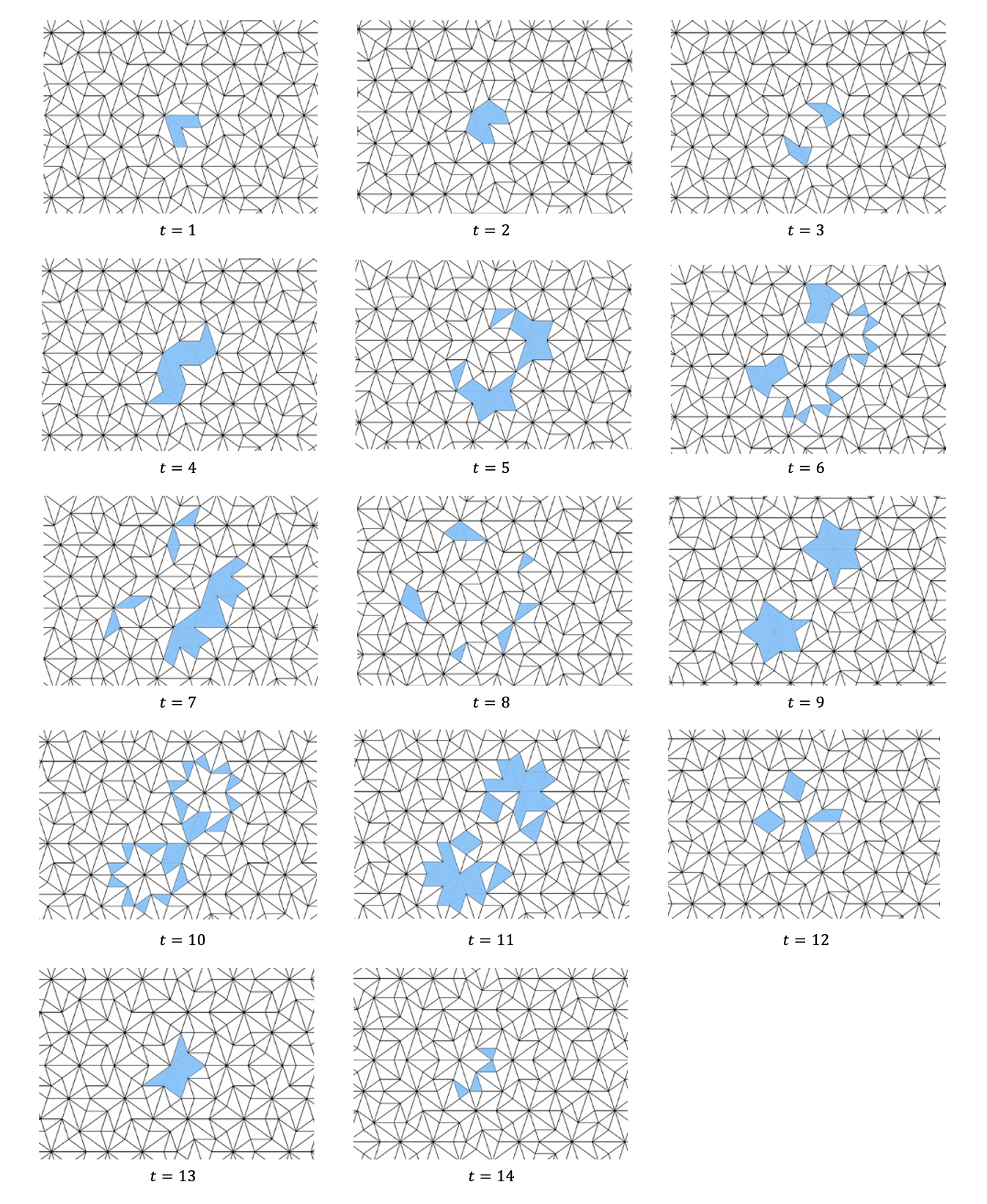}
\caption{A period-14 oscillator.}\label{fig:oscillator}
\end{figure}

\begin{question}
Can one classify oscillators in the Robinson triangle Penrose Game of Life?
\end{question}
In the course of this project, we discovered a period-14 oscillator, shown in Figure~\ref{fig:oscillator}. Are there any more? Does there exist a natural classification scheme for these oscillators?

\section*{Acknowledgments}
This project was begun with support by the J. Reid and Polly Anderson Endowment at Denison University. The authors thank the anonymous referee for their insightful comments.

\providecommand{\bysame}{\leavevmode\hbox to3em{\hrulefill}\thinspace}
\bibliographystyle{amsplain}

\begin{landscape}
\begin{table}
\begin{tabular}{r|cccccccc}
&$d-d$ & $d-k$ & $d-N$ & $d-r$ & $d-R$ & $d-t$ & $d-w-w$ & $k-k$  \\ \hline
$N_{10}^1$ &        &     &       &       & 1     &         &         &       \\ \hline
$N_{12}^1$ &        &     &       & 2     & 3     & 2       &         &       \\
$N_{12}^2$ &        &     &       & 2     & 3     & 2       &         &       \\  \hline
$N_{14}^1$ & 5       &     &       &       & 16    &         & 50      &       \\
$N_{14}^2$ & 1       & 3   & 4     &       & 6     &         & 10      &       \\
$N_{14}^3$ &        & 2   & 2     &       & 1     &         &         & 1     \\
$N_{14}^4$ &        & 2   & 2     &       & 1     &         &         & 1     \\  \hline
$N_{16}^1$ &1       & 3   & 4     &       & 7     &         & 10      &       \\
$N_{16}^2$ &1       & 3   & 4     &       & 7     &         & 10      &       \\
$N_{16}^3$ &5       &     &       &       & 16    &         & 50      &       \\
$N_{16}^4$ &1       & 3   & 4     & 2     & 8     & 2       & 10      &       \\
$N_{16}^5$ &        & 2   & 2     & 2     & 3     & 2       &         & 1     \\  \hline
$N_{18}^1$ & 5       & 2   & 2     &       & 16    &         & 50      & 1     \\
$N_{18}^2$ &1       & 3   & 4     & 2     & 9     & 2       & 10      &       \\
$N_{18}^3$ &        & 2   & 2     & 2     & 4     & 2       &         & 1     \\
\end{tabular}
\caption{The number of still lifes of the form $d-d$ through $k-k$ found in all neighborhoods.}\label{tab1}
\end{table}
\begin{table}
\begin{tabular}{r|cccccccc}
&$k-N$     & $k-R$ & $k-w-w$ & $d-n-w$ & $k-n-w$ & $N-N$     & $d-n-n$   & $n-n-R$ \\ \hline
$N_{10}^1$ &        & 1   &       &       &       &         &         &       \\ \hline
$N_{12}^1$ &        &     &       & 6     &       &         &         &       \\
$N_{12}^2$ &        &     &       & 6     &       &         &         &       \\ \hline
$N_{14}^1$ &       &     &       &       &       &         &         &       \\
$N_{14}^2$ &        & 3   & 10    & 20    &       &         &         &       \\
$N_{14}^3$ &4       & 3   & 8     & 6     & 12    & 3       & 4       & 5     \\
$N_{14}^4$ &4       & 3   & 8     & 6     & 12    & 3       & 4       & 5     \\ \hline
$N_{16}^1$ &        & 3   & 10    & 20    &       &         &         &       \\
$N_{16}^2$ &        & 3   & 10    & 20    &       &         &         &       \\
$N_{16}^3$ &        & 1   &       &       &       &         &         &       \\
$N_{16}^4$ &        & 2   & 10    & 26    &       &         &         &       \\
$N_{16}^5$ &4       & 2   & 8     & 12    & 12    & 3       & 4       & 5     \\ \hline
$N_{18}^1$ & 4       & 2   & 8     & 6     & 12    & 3       & 4       & 5     \\
$N_{18}^2$ &        & 2   & 10    & 26    &       &         &         &       \\
$N_{18}^3$ &4       & 2   & 8     & 12    & 12    & 3       & 4       & 4     \\
\end{tabular}
\caption{The number of still lifes of the form $d-N$ through $n-n-R$ found in all neighborhoods.}\label{tab2}
\end{table}
\begin{table}
\begin{tabular}{r|cccccccc}
&$n-n-w-w$ & $N-R$ & $N-t$   & $n-N-w$ & $n-R-w$ & $n-w-w-w$ & $r-R$     & $R-R$   \\ \hline
$N_{10}^1$ &        &     & 2     &       &       & 1       &         &       \\ \hline
$N_{12}^1$ &       &     & 2     &       & 6     & 3       & 1       & 1     \\
$N_{12}^2$ &1       &     & 2     &       & 6     & 2       & 1       & 1     \\ \hline
$N_{14}^1$ &1       &     & 2     &       &       &         &         & 5     \\
$N_{14}^2$ &1       & 6   &       &       & 20    & 20      &         & 3     \\
$N_{14}^3$ &16      & 8   &       & 26    & 18    & 10      &         & 2     \\
$N_{14}^4$ &16      & 8   &       & 26    & 18    & 10      &         & 2     \\ \hline
$N_{16}^1$ &2       & 6   & 2     &       & 20    & 21      &         & 3     \\
$N_{16}^2$ &2       & 6   & 2     &       & 20    & 21      &         & 3     \\
$N_{16}^3$ &2       &     & 2     &       &       & 1       &         & 5     \\
$N_{16}^4$ &1       & 6   &       &       & 26    & 22      & 1       & 4     \\
$N_{16}^5$ &16      & 8   &       & 26    & 24    & 12      & 1       & 3     \\ \hline
$N_{18}^1$ & 16      & 8   &       & 26    & 18    & 10      &         & 7     \\
$N_{18}^2$ &2       & 6   & 2     &       & 26    & 23      & 1       & 4     \\
$N_{18}^3$ &17      & 8   & 2     & 26    & 25    & 13      & 1       & 3     \\
\end{tabular}
\caption{The number of still lifes of the form $n-n-w-w$ through $R-R$ found in all neighborhoods.}\label{tab3}
\end{table}
\begin{table}
\begin{tabular}{r|cccccccc}
&$r-t$     & $R-t$ & $N-w-w$ & $r-w-w$ & $R-w-w$ & $t-w-w$   & $w-w-w-w$ & Total \\ \hline
$N_{10}^1$ &2       &     &       &       &       &         &         & 7     \\ \hline
$N_{12}^1$ &        & 4   &       & 3     & 3     & 6       &         & 42    \\
$N_{12}^2$ &        & 4   &       & 3     & 3     & 6       &         & 42    \\ \hline
$N_{14}^1$ &       &     &       &       & 50    &         & 25      & 154   \\
$N_{14}^2$ & 2       &     & 20    &       & 20    &         & 5       & 154   \\
$N_{14}^3$ &2       &     & 14    &       & 7     &         &         & 154   \\
$N_{14}^4$ &2       &     & 14    &       & 7     &         &         & 154   \\ \hline
$N_{16}^1$ &2       &     & 20    &       & 20    &         & 5       & 159   \\
$N_{16}^2$ &2       &     & 20    &       & 20    &         & 5       & 159   \\
$N_{16}^3$ &2       &     &       &       & 50    &         & 25      & 159   \\
$N_{16}^4$ &        & 4   & 20    & 3     & 23    & 6       & 5       & 189   \\
$N_{16}^5$ &        & 4   & 14    & 3     & 10    & 6       &         & 189   \\ \hline
$N_{18}^1$ &       &     & 14    &       & 57    &         & 25      & 301   \\
$N_{18}^2$ &        & 4   & 20    & 3     & 23    & 6       & 5       & 194   \\
$N_{18}^3$ &        & 4   & 14    & 3     & 10    & 6       &         & 194  
\end{tabular}
\caption{The number of still lifes of the form $r-t$ through $w-w-w-w$ found in all neighborhoods, along with the total.}\label{tab4}
\end{table}
\end{landscape}

\end{document}